\documentclass[aip,preprint]{revtex4-1}
\usepackage{graphicx}
\begin{document}
\title{Formation and anisotropic magnetoresistance of Co/Pt nano-contacts through aluminum oxide barrier} 
\author{Muftah Al-Mahdawi}
\email{mahdawi@ecei.tohoku.ac.jp}
\author{Masashi Sahashi}
\affiliation{Department of Electronic Engineering, Tohoku University, Sendai 980-8579, Japan}
\date{\today}
\begin{abstract}
We report on the observation of anisotropic magnetoresistance (AMR) in vertical asymmetric nano-contacts (NCs) made through AlO$_x$ nano-oxide layer (NOL) formed by ion-assisted oxidation method in the film stack of Co/AlO$_x$-NOL/Pt. Analysis of NC formation was based on \emph{in situ} conductive atomic force microscopy and transmission electron microscopy. Depending on the purity of NCs from Al contamination, we observed up to 29\% AMR ratio at room temperature.
\end{abstract}
\maketitle
\begin{figure}
	\caption{(a) cAFM current image of AlO$_x$-NOL over Co thin-film taken. Sample line profiles shown inset. (b) Superposition of current image (in light blue) over topography profile. (c) Number and occupancy of conductive paths vs.~thresholding current level. A plateau is seen at 1.6--1.8 nA. (d) Distribution of $V_{bi}$ measured from I-V curves of a semiconducting tip over single-film Co and Al, and Co/AlO$_x$-NOL conductive paths. Conductive paths divided into two groups: Co-like and Al-like. (e) TEM image of Co/AlO$_x$-NOL/Pt showed direct connections between Pt and Co layers through AlO$_x$.}
	\label{fig:cAFM_image}
	\includegraphics[width=1\textwidth]{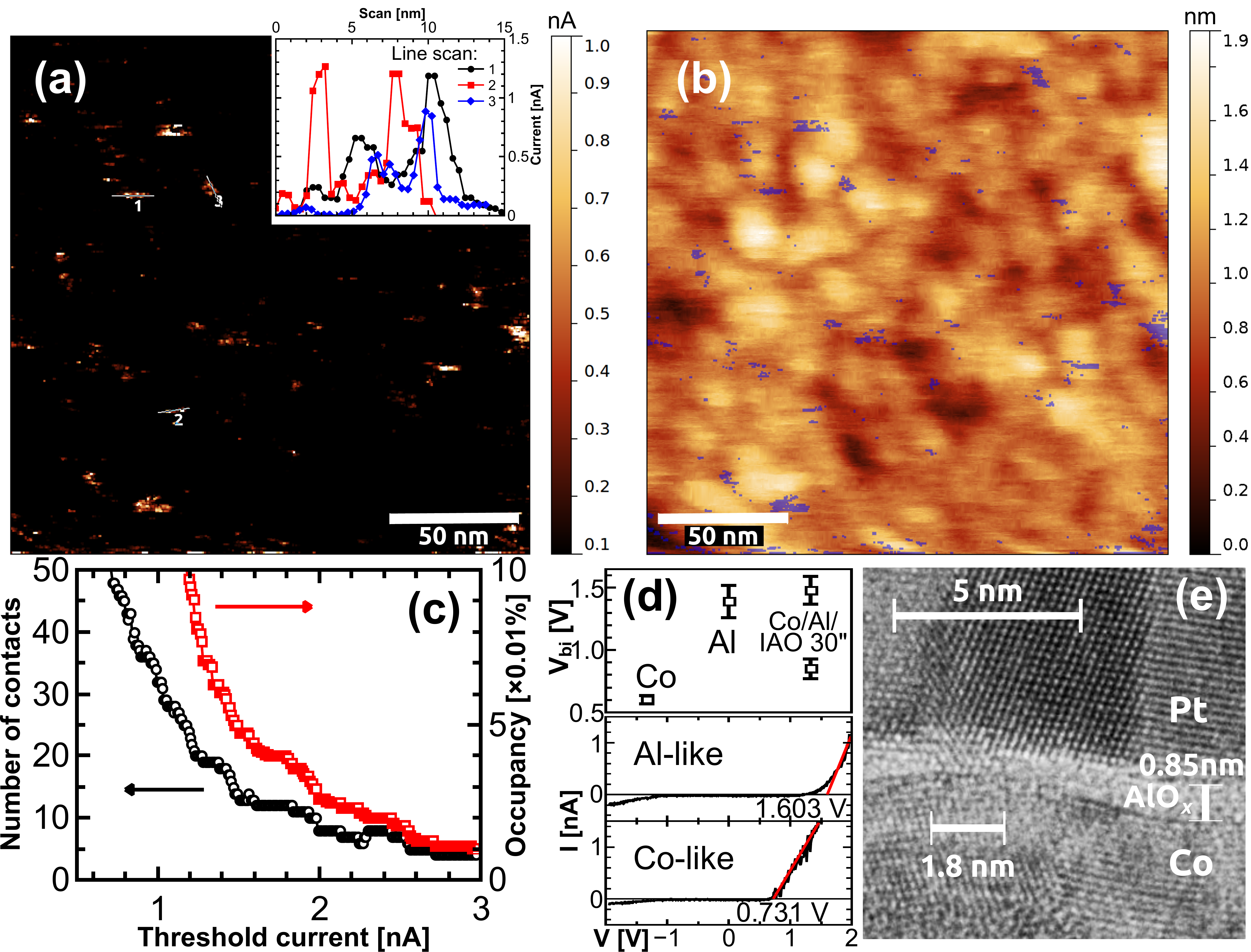}
\end{figure}
\begin{figure}
    \caption{(a) Pillar A (180 nm in diameter) with Al-dominant NCs had a very low AMR ratio at 300 K which increased at 5 K. (b) For pillar B (200 nm in diameter) with Pt-dominant NCs, AMR ratio increased substantially at 300 K but decreased at 5 K with widened flattening around 90$^\circ$ and 270$^\circ$ indicating increased in-plane anisotropy (insets are dependence of AMR ratio on applied field). (c) AMR ratio had an anomaly at 150--160K, which can be attributed to onset of coupling with CoO. (d) A schematic showing the presence of both Al-NCs and Pt-NCs (the arrow refers to induced magnetization in Pt-NCs but the direction does not represent the actual configuration), with CoO present beneath AlO$_x$.}
	\label{fig:AMR}
	\includegraphics[width=1\textwidth]{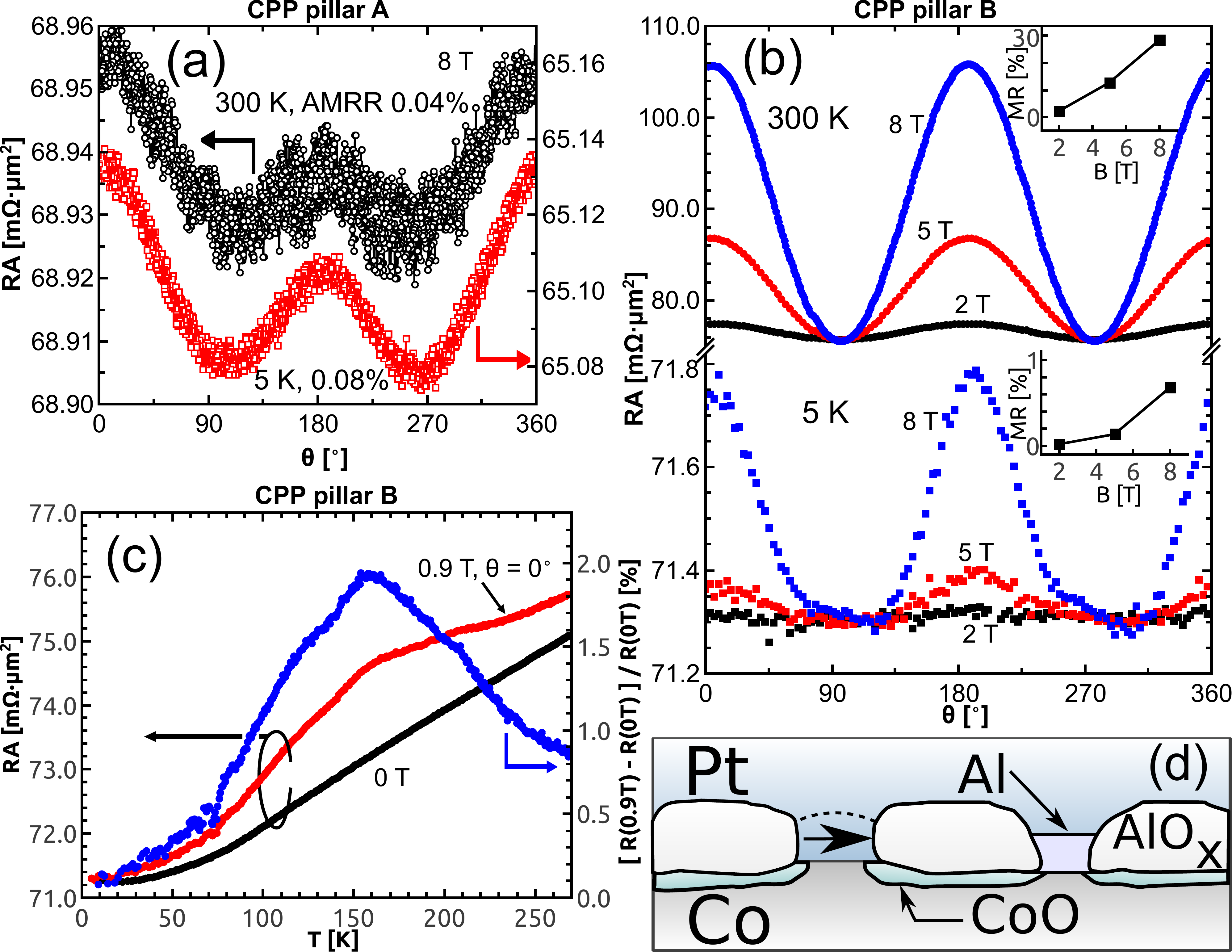}
	
\end{figure}

Ballistic transport either through a tunneling barrier or nano-contacts (NCs), where contact diameter is much smaller than electron's mean free path, offers a lot of information about electron's local Density Of States (DOS) near Fermi level. In systems with Spin-Orbit Coupling (SOC), local DOS dependence on local magnetization direction leads to the effects of tunneling anisotropic magnetoresistance (TAMR), for tunneling conduction,\cite{gould2004,park2008} and nano-contact AMR (NC-AMR), in the case of ballistic transport through nanometer-sized metallic contacts.\cite{bolotin2006} In principle, effect of DOS anisotropy should be observable up to room temperature\cite{shick2006} but it was not, either due to randomly trapped charges at the barrier,\cite{park2008} or due to instability of atoms in planar NCs.\cite{bolotin2006} Vertical NCs between films provide a more stable alternative to planar NCs, but fabrication of NCs in this geometry is difficult with current lithography techniques.\cite{ralls1989} Bombarding the surface of a thin Al layer with low-energy (65 eV) Ar$^+$-ions in the presence of oxygen atmosphere ($\approx 1 \times 10^{-2}$Pa partial pressure), the so-called Ion-Assisted Oxidation (IAO) method, can be used to form an AlO$_x$ Nano-Oxide-Layer (NOL) with multiple holes 1--2 nm in diameter.\cite{takagishi2009} IAO method has been used to make ferromagnetic NCs between the upper and bottom layers; if short oxidation times are used (20--40 sec).\cite{fuke2007, shiokawa2011} IAO method is relatively easy compared to lithography and the resulting NCs are stable even at elevated temperatures, but the formation mechanism is inherently a random bottom-up process.\\
By placing an interface (Co/Pt) with a strong SOC adjacent to a tunneling barrier, Ref.~\onlinecite{park2008} could increase TAMR ratio two-folds. In this paper, we report on the formation of Co/Pt asymmetric NCs through AlO$_x$-NOL, and the resulting AMR behavior.\\
To study the formation of AlO$_x$-NOL over Co layer by IAO, we used \emph{in situ} atomic force microscopy in conductive mode (cAFM) with $10^{-8}$-Pa base pressure to characterize the surface topography and current profile of AlO$_x$-NOL. A Si tip coated with heavily-doped diamond (CDT-CONTR) from NANOSENSORS\texttrademark~was used. The film design was (thickness in nm): thermally-oxidized silicon substrate/electrode layer(Ta (5)/Cu (285)/Ta (30)/chemical-mechanical polishing)/Ta (10)/Ru (2)/Co (2)/Al (1.3)/IAO 30 sec exposure time. After deposition, the sample was transferred in vacuum to cAFM chamber. Image analysis was done by extension scripts to Gwyddion software.\cite{necas2011} Electrode layer/Ta (10)/Ru (2)/Co (2)/Al (1.3)/IAO 30 sec/Pt (7) was used for transport measurement. Co, Pt, Ru, and Ta were deposited by dc magnetron sputtering. Ion-beam sputtering was used for Al deposition, with an assist-ion-gun used for IAO at the same chamber. All chambers are connected through an ultra-high-vacuum ($10^{-6}$-Pa base pressure) transfer tube. Current-In-Plane-Tunneling (CIPT) method was used to measure Resistance-Area (RA) product of unpatterned films. Current-Perpendicular-to-Plane (CPP) geometry pillars with circular cross-section (180--400 nm in diameter) were patterned by Ar$^+$ ion-milling and lithography techniques, where RA was also deduced from the slope of CPP pillars dc resistance vs.~area inverse (1/A) plot. Cross-sectional high-resolution transmission electron micrographs (TEM) were obtained as an additional indication of the shape and material composition of NCs.
Zero-bias differential resistance of the NCs ($R_0$) dependence on magnetization direction was measured by applying a large magnetic field and varying the angle ($\theta$) between normal to film plane and the applied field, at 300-K and 5-K temperatures. Also $R_0$ dependence on temperature in the heating direction was measured for 0-T and 0.9-T magnetic fields applied out-of-plane, after cooling from 350 K at the same magnetic field used in measurement. Angles of $0^\circ$ and $90^\circ$ corresponds to out-of-plane and in-plane applied field directions, respectively. AMR ratio is defined as $(R_0(\theta)-R_0(0^\circ))/R_0(0^\circ)$.\\
cAFM current and topography scan images of 200 nm $\times$ 200 nm area at 1.8-V tip-to-sample bias are shown in Figs.~\ref{fig:cAFM_image}(a),(b). Several high-current paths were observed [sample profiles are in the inset of Fig.~\ref{fig:cAFM_image}(a)], which correlate with the valleys between the grains of AlO$_x$-NOL [Fig.~\ref{fig:cAFM_image}(b)]. This indicates that holes form at the grain boundaries during oxidation process, in agreement with previous cAFM study of AlO$_x$-NOL over Fe$_{50}$Co$_{50}$,\cite{shiokawa2011} and that the top ferromagnetic layer determines MR performance in Fe$_{x}$Co$_{1-x}$/AlO$_x$-NOL/Fe$_{y}$Co$_{1-y}$ structures.\cite{kishi2012}
To estimate the number of conductive paths and mean diameter, current image was turned into a binary image by thresholding; then number, occupancy, and mean diameter of paths were computed for each current threshold level [Fig.~\ref{fig:cAFM_image}(c)]. A dilation-erosion filter was used to eliminate single-pixel noise in current image beforehand. The observed conductive paths at higher threshold levels are due to metallic contacts, whereas at lower levels noise from tunneling through AlO$_x$ prevails.\cite{kishi2012} A plateau in number of contacts and occupancy vs.~threshold current at 1.6--1.8 nA, attributed to a transition in counting from metallic contacts to tunneling noise, can be used to deduce characteristics of NCs. Number density, area occupancy, and mean diameter are 300 $\mathrm{\mu m^{-2}}$, 0.04\%, and 1.25 nm, respectively. 
Semiconducting tips over metal films form a Schottky barrier, with junction built-in potential ($V_{bi}$) being the difference between work functions of tip material and sample surface metal. The tip was calibrated by measuring the I-V curves over Al and Co single 20-nm films, and rectifying I-V characteristic was confirmed. Average $V_{bi}$ values were found to differ by 0.79 V between Co and Al. This corresponds well with the difference between work-functions of Al and Co ($\approx$0.8 eV). Using the same tip, I-V curves were measured over multiple conductive paths of Co/AlO$_x$-NOL. The values of $V_{bi}$ showed separation into two groups, one group had values closer to Co and the other closer to Al [Fig.~\ref{fig:cAFM_image}(d)]. We imply that the conductive paths are metals with a random distribution of Co and Al. Thus, by capping Co/AlO$_x$-NOL with Pt, there will be a random distribution of Co/Al/Pt and Co/Pt nano-contacts. TEM images of Co/AlO$_x$-NOL/Pt showed direct connections between Pt and Co layers through AlO$_x$ at few places [Fig.~\ref{fig:cAFM_image}(e)]. The crystal structure and orientation of NC region is same as top Pt layer and different from Co under-layer. In support of expectations from cAFM results, Pt filled most of the contact volume with a similar NC diameter of 1.8 nm. Due to low atomic numbers of Al and O, it was not possible to distinguish Al from AlO$_x$ in TEM image. AlO$_x$ thickness ranged 0.75--0.9 nm which is thinner than the 1.3-nm expected thickness of Al. The effective electrical thickness of AlO$_x$ was 0.8 nm, determined from BDR model fitting\cite{brinkman1970} to I-V curves of long-time oxidized junctions (IAO 180 sec). The reason for this discrepancy is due to further consideration.\\
RA products of Co/AlO$_x$-NOL/Pt films from measurement of CIPT and CPP-pillars resistance were 100--150 and 81 $\mathrm{m\Omega\cdot \mu m^2}$, respectively. Also, resistance increased with increasing temperature and bias voltage (not shown), indicating metallic contact formation through AlO$_x$-NOL. This low RA, compared with FeCo-NCs,\cite{fuke2007} is from effect of Pt on AlO$_x$-NOL. When Pt capping was replaced with Co, RA increased to 330 $\mathrm{m\Omega\cdot \mu m^2}$, and resistance increased with decreasing temperature. Pt is not inert in the presence of Co and AlO$_x$,\cite{lu1995-a} the effect of Pt on NCs would be to reduce the oxygen impurities and enhance metallic contact formation.\\
CPP pillars from same substrate showed characteristics that could be grouped into two groups, from which the best cases are reported next. We attribute this grouping to prevalence of either Al or Pt NCs in the area where the CPP pillar was patterned. In pillar A, $R_0-\theta$ curve had a low AMR ratio of 0.04\%~at 300 K, which increased to 0.08\%~ at 5 K under 8-T magnetic field [Fig.~\ref{fig:AMR}(a)]. Contrarily, pillar B had up to 28.6\%~AMR ratio at 300 K [Fig.~\ref{fig:AMR}(b)]. The AMR ratio of pillar B increased with increasing applied field [insets of Fig.~\ref{fig:AMR}(b)] and decreased with increasing temperature at high temperatures (T $>$ 170 K) [Fig.~\ref{fig:AMR}(c)]. The qualitative proportionality of AMR ratio to \emph{B/T} can be attributed to Curie-Weiss-like dependency of induced magnetization in Pt\cite{park2008} at NCs. At the contact area between Pt and Co, the induced magnetization can show a superparamegnetic behavior. Possible mechanism for superparamagnetism is the presence of fluctuating moments in Pt-NCs exchange coupled to Co layer,\cite{celinski1990} or diffused Co atoms,\citep{crangle1965} in addition to reduced ``magnetic particle'' size at NC. We attribute the large AMR ratio of pillar B to the large DOS anisotropy of Co/Pt interface.\cite{park2008} In the case of NC geometry, the reduced dimensionality increases the magnitude of SOC,\cite{doudin2008} which would increase the DOS anisotropy more, compared to extended electrodes. We expect that the presence of the weak-SOC Al contamination at NCs lowers the DOS anisotropy and introduces shunting paths against purer Co/Pt NCs, thus lowering AMR ratio in the group of pillar A [Fig.~\ref{fig:AMR}(d)].\\
At 150--160 K, pillar B (and other pillars having large AMR ratio $>$ 0.5\%) had an anomaly in AMR ratio, with a decreasing AMR ratio at low temperature [Fig.~\ref{fig:AMR}(c)], whereas for pillar A the low AMR ratio limited the ability to resolve such an anomaly. The $R_0-\theta$ curve at 5 K and 8 T has a widened flattening around 90$^\circ$ and 270$^\circ$. This indicates an induced in-plane anisotropy preventing the NC magnetic moments from aligning with applied field leading to an apparent additional fourfold-symmetry dependence and suppressing the AMR ratio down to 0.7\% at 5 K[Fig.~\ref{fig:AMR}(b)]. Due to small magnitude of Pt induced magnetization (0.2--0.6 $\mu_B$),\cite{geissler2001,*suzuki2005} even a small induced anisotropy results in a large anisotropy field. The temperature of this anomaly coincides well with the blocking temperature ($T_B$) of CoO in NOL systems.\cite{sahashi2007} Formation of CoO (or Co-O bonds) beneath AlO$_x$ and around NCs during oxidation process is within expectation [Fig.~\ref{fig:AMR}(d)]. We propose that below $T_B$ the exchange coupling between Pt and CoO induces such an anisotropy. The amplitude and nature of coupling is difficult to determine in this report due to unknown magnetization of Pt-NCs and coupling of Pt moments to both Co and CoO, which can lead to non-trivial coupling configurations.\cite{slonczewski1995,*fukuzawa2002}\\
The observed AMR cannot be attributed to the bulk AMR of 2-nm Co layer, as it would amount to a 0.003\%~resistance change in the 81-$\mathrm{m\Omega\mu m^2}$ pillar RA (Assuming $\mathrm{RA_{Co}=124.8~\mu\Omega\mu m^2}$, and $\mathrm{AMRR_{Co}=1.7\%}$) and no AMR could be measured in CPP pillars of Ru/Co 2 nm/Ru. To check the effect of removing NCs on AMR behavior, thicker AlO$_x$ layer with longer IAO time was used to separate Co and Pt layers completely (Co 2/Al 1.3/IAO 180 sec/Al 1.3/IAO 240 sec/Pt 7, thickness in nm). CPP-RA was 10 $\mathrm{k\Omega\cdot \mu m^2}$, and $R_0-\mathrm{\theta}$ curve showed similar behavior ($R(0^\circ) < R(90^\circ)$) to sample B (Co/AlO$_x$/Pt) of Ref.~\onlinecite{park2008} with 0.06\% tunneling AMR ratio at 3K.\\
In conclusion, we characterized the formation of randomly distributed Al-NCs or Pt-NCs with an interface to Co through AlO$_x$-NOL, using the bottom-up process of IAO. Because of large anisotropic DOS of Pt/Co system, we were able to observe up to 29\% NC-AMR ratio at room temperature. Due to randomness in NC formation mechanism through AlO$_x$-NOL, mixed results were obtained from CPP pillars made of the same film because of the presence of Al contamination. The expected presence of CoO ($T_B$ = 150 K in NOL structure) resulted in an induced in-plane anisotropy, suppressing AMR ratio at low temperature. More clarification is needed for the role of coupling modulation in Co-CoO/Pt NC system. The non-scattering transport of electrons through NCs allows for effective application of high electric fields within the metal, the prospects of which are under investigation. The thermal stability of vertical NCs, adding the relative ease of making NCs by IAO method, promise real-world applications, but one major hurdle is improving the control of NCs formation with a higher yield, which is an ongoing research effort.
\begin{acknowledgments}
 M. A. acknowledges financial support from Ministry of Education, Culture, Sports, Science and Technology (MEXT). Authors thank Dr.~Y.~Shiokawa and Dr.~T.~Nozaki for their help and comments.
\end{acknowledgments}

%


\begin{thebibliography}{20}%
\makeatletter
\providecommand \@ifxundefined [1]{%
 \@ifx{#1\undefined}
}%
\providecommand \@ifnum [1]{%
 \ifnum #1\expandafter \@firstoftwo
 \else \expandafter \@secondoftwo
 \fi
}%
\providecommand \@ifx [1]{%
 \ifx #1\expandafter \@firstoftwo
 \else \expandafter \@secondoftwo
 \fi
}%
\providecommand \natexlab [1]{#1}%
\providecommand \enquote  [1]{``#1''}%
\providecommand \bibnamefont  [1]{#1}%
\providecommand \bibfnamefont [1]{#1}%
\providecommand \citenamefont [1]{#1}%
\providecommand \href@noop [0]{\@secondoftwo}%
\providecommand \href [0]{\begingroup \@sanitize@url \@href}%
\providecommand \@href[1]{\@@startlink{#1}\@@href}%
\providecommand \@@href[1]{\endgroup#1\@@endlink}%
\providecommand \@sanitize@url [0]{\catcode `\\12\catcode `\$12\catcode
  `\&12\catcode `\#12\catcode `\^12\catcode `\_12\catcode `\%12\relax}%
\providecommand \@@startlink[1]{}%
\providecommand \@@endlink[0]{}%
\providecommand \url  [0]{\begingroup\@sanitize@url \@url }%
\providecommand \@url [1]{\endgroup\@href {#1}{\urlprefix }}%
\providecommand \urlprefix  [0]{URL }%
\providecommand \Eprint [0]{\href }%
\providecommand \doibase [0]{http://dx.doi.org/}%
\providecommand \selectlanguage [0]{\@gobble}%
\providecommand \bibinfo  [0]{\@secondoftwo}%
\providecommand \bibfield  [0]{\@secondoftwo}%
\providecommand \translation [1]{[#1]}%
\providecommand \BibitemOpen [0]{}%
\providecommand \bibitemStop [0]{}%
\providecommand \bibitemNoStop [0]{.\EOS\space}%
\providecommand \EOS [0]{\spacefactor3000\relax}%
\providecommand \BibitemShut  [1]{\csname bibitem#1\endcsname}%
\let\auto@bib@innerbib\@empty
\bibitem [{\citenamefont {Gould}\ \emph {et~al.}(2004)\citenamefont {Gould},
  \citenamefont {R\"{u}ster}, \citenamefont {Jungwirth}, \citenamefont
  {Girgis}, \citenamefont {Schott}, \citenamefont {Giraud}, \citenamefont
  {Brunner}, \citenamefont {Schmidt},\ and\ \citenamefont
  {Molenkamp}}]{gould2004}%
  \BibitemOpen
  \bibfield  {author} {\bibinfo {author} {\bibfnamefont {C.}~\bibnamefont
  {Gould}}, \bibinfo {author} {\bibfnamefont {C.}~\bibnamefont {R\"{u}ster}},
  \bibinfo {author} {\bibfnamefont {T.}~\bibnamefont {Jungwirth}}, \bibinfo
  {author} {\bibfnamefont {E.}~\bibnamefont {Girgis}}, \bibinfo {author}
  {\bibfnamefont {G.~M.}\ \bibnamefont {Schott}}, \bibinfo {author}
  {\bibfnamefont {R.}~\bibnamefont {Giraud}}, \bibinfo {author} {\bibfnamefont
  {K.}~\bibnamefont {Brunner}}, \bibinfo {author} {\bibfnamefont
  {G.}~\bibnamefont {Schmidt}}, \ and\ \bibinfo {author} {\bibfnamefont
  {L.~W.}\ \bibnamefont {Molenkamp}},\ }\href {\doibase
  10.1103/PhysRevLett.93.117203} {\bibfield  {journal} {\bibinfo  {journal}
  {Phys. Rev. Lett.}\ }\textbf {\bibinfo {volume} {93}},\ \bibinfo {pages}
  {117203} (\bibinfo {year} {2004})}\BibitemShut {NoStop}%
\bibitem [{\citenamefont {Park}\ \emph {et~al.}(2008)\citenamefont {Park},
  \citenamefont {Wunderlich}, \citenamefont {Williams}, \citenamefont {Joo},
  \citenamefont {Jung}, \citenamefont {Shin}, \citenamefont {Olejn\'{i}k},
  \citenamefont {Shick},\ and\ \citenamefont {Jungwirth}}]{park2008}%
  \BibitemOpen
  \bibfield  {author} {\bibinfo {author} {\bibfnamefont {B.~G.}\ \bibnamefont
  {Park}}, \bibinfo {author} {\bibfnamefont {J.}~\bibnamefont {Wunderlich}},
  \bibinfo {author} {\bibfnamefont {D.~A.}\ \bibnamefont {Williams}}, \bibinfo
  {author} {\bibfnamefont {S.~J.}\ \bibnamefont {Joo}}, \bibinfo {author}
  {\bibfnamefont {K.~Y.}\ \bibnamefont {Jung}}, \bibinfo {author}
  {\bibfnamefont {K.~H.}\ \bibnamefont {Shin}}, \bibinfo {author}
  {\bibfnamefont {K.}~\bibnamefont {Olejn\'{i}k}}, \bibinfo {author}
  {\bibfnamefont {A.~B.}\ \bibnamefont {Shick}}, \ and\ \bibinfo {author}
  {\bibfnamefont {T.}~\bibnamefont {Jungwirth}},\ }\href {\doibase
  10.1103/PhysRevLett.100.087204} {\bibfield  {journal} {\bibinfo  {journal}
  {Phys. Rev. Lett.}\ }\textbf {\bibinfo {volume} {100}},\ \bibinfo {pages}
  {087204} (\bibinfo {year} {2008})}\BibitemShut {NoStop}%
\bibitem [{\citenamefont {Bolotin}\ \emph {et~al.}(2006)\citenamefont
  {Bolotin}, \citenamefont {Kuemmeth}, \citenamefont {Pasupathy},\ and\
  \citenamefont {Ralph}}]{bolotin2006}%
  \BibitemOpen
  \bibfield  {author} {\bibinfo {author} {\bibfnamefont {K.~I.}\ \bibnamefont
  {Bolotin}}, \bibinfo {author} {\bibfnamefont {F.}~\bibnamefont {Kuemmeth}},
  \bibinfo {author} {\bibfnamefont {A.~N.}\ \bibnamefont {Pasupathy}}, \ and\
  \bibinfo {author} {\bibfnamefont {D.~C.}\ \bibnamefont {Ralph}},\ }\href
  {\doibase 10.1021/nl0522936} {\bibfield  {journal} {\bibinfo  {journal} {Nano
  Lett.}\ }\textbf {\bibinfo {volume} {6}},\ \bibinfo {pages} {123} (\bibinfo
  {year} {2006})}\BibitemShut {NoStop}%
\bibitem [{\citenamefont {Shick}\ \emph {et~al.}(2006)\citenamefont {Shick},
  \citenamefont {M\'aca}, \citenamefont {Ma\v{s}ek},\ and\ \citenamefont
  {Jungwirth}}]{shick2006}%
  \BibitemOpen
  \bibfield  {author} {\bibinfo {author} {\bibfnamefont {A.~B.}\ \bibnamefont
  {Shick}}, \bibinfo {author} {\bibfnamefont {F.}~\bibnamefont {M\'aca}},
  \bibinfo {author} {\bibfnamefont {J.}~\bibnamefont {Ma\v{s}ek}}, \ and\
  \bibinfo {author} {\bibfnamefont {T.}~\bibnamefont {Jungwirth}},\ }\href
  {\doibase 10.1103/PhysRevB.73.024418} {\bibfield  {journal} {\bibinfo
  {journal} {Phys. Rev. B}\ }\textbf {\bibinfo {volume} {73}},\ \bibinfo
  {pages} {024418} (\bibinfo {year} {2006})}\BibitemShut {NoStop}%
\bibitem [{\citenamefont {Ralls}, \citenamefont {Buhrman},\ and\ \citenamefont
  {Tiberio}(1989)}]{ralls1989}%
  \BibitemOpen
  \bibfield  {author} {\bibinfo {author} {\bibfnamefont {K.~S.}\ \bibnamefont
  {Ralls}}, \bibinfo {author} {\bibfnamefont {R.~A.}\ \bibnamefont {Buhrman}},
  \ and\ \bibinfo {author} {\bibfnamefont {R.~C.}\ \bibnamefont {Tiberio}},\
  }\href {\doibase http://dx.doi.org/10.1063/1.102001} {\bibfield  {journal}
  {\bibinfo  {journal} {Appl. Phys. Lett.}\ }\textbf {\bibinfo {volume} {55}},\
  \bibinfo {pages} {2459} (\bibinfo {year} {1989})}\BibitemShut {NoStop}%
\bibitem [{\citenamefont {Takagishi}\ \emph {et~al.}(2009)\citenamefont
  {Takagishi}, \citenamefont {Fuke}, \citenamefont {Hashimoto}, \citenamefont
  {Iwasaki}, \citenamefont {Kawasaki}, \citenamefont {Shiozaki},\ and\
  \citenamefont {Sahashi}}]{takagishi2009}%
  \BibitemOpen
  \bibfield  {author} {\bibinfo {author} {\bibfnamefont {M.}~\bibnamefont
  {Takagishi}}, \bibinfo {author} {\bibfnamefont {H.~N.}\ \bibnamefont {Fuke}},
  \bibinfo {author} {\bibfnamefont {S.}~\bibnamefont {Hashimoto}}, \bibinfo
  {author} {\bibfnamefont {H.}~\bibnamefont {Iwasaki}}, \bibinfo {author}
  {\bibfnamefont {S.}~\bibnamefont {Kawasaki}}, \bibinfo {author}
  {\bibfnamefont {R.}~\bibnamefont {Shiozaki}}, \ and\ \bibinfo {author}
  {\bibfnamefont {M.}~\bibnamefont {Sahashi}},\ }\href {\doibase
  http://dx.doi.org/10.1063/1.3073952} {\bibfield  {journal} {\bibinfo
  {journal} {J. Appl. Phys.}\ }\textbf {\bibinfo {volume} {105}},\ \bibinfo
  {eid} {07B725} (\bibinfo {year} {2009})}\BibitemShut {NoStop}%
\bibitem [{\citenamefont {Fuke}\ \emph {et~al.}(2007)\citenamefont {Fuke},
  \citenamefont {Hashimoto}, \citenamefont {Takagishi}, \citenamefont
  {Iwasaki}, \citenamefont {Kawasaki}, \citenamefont {Miyake},\ and\
  \citenamefont {Sahashi}}]{fuke2007}%
  \BibitemOpen
  \bibfield  {author} {\bibinfo {author} {\bibfnamefont {H.~N.}\ \bibnamefont
  {Fuke}}, \bibinfo {author} {\bibfnamefont {S.}~\bibnamefont {Hashimoto}},
  \bibinfo {author} {\bibfnamefont {M.}~\bibnamefont {Takagishi}}, \bibinfo
  {author} {\bibfnamefont {H.}~\bibnamefont {Iwasaki}}, \bibinfo {author}
  {\bibfnamefont {S.}~\bibnamefont {Kawasaki}}, \bibinfo {author}
  {\bibfnamefont {K.}~\bibnamefont {Miyake}}, \ and\ \bibinfo {author}
  {\bibfnamefont {M.}~\bibnamefont {Sahashi}},\ }\href {\doibase
  10.1109/TMAG.2007.893117} {\bibfield  {journal} {\bibinfo  {journal} {IEEE
  Trans. Magn.}\ }\textbf {\bibinfo {volume} {43}},\ \bibinfo {pages} {2848}
  (\bibinfo {year} {2007})}\BibitemShut {NoStop}%
\bibitem [{\citenamefont {Shiokawa}\ \emph {et~al.}(2011)\citenamefont
  {Shiokawa}, \citenamefont {Shiota}, \citenamefont {Watanabe}, \citenamefont
  {Otsuka}, \citenamefont {Doi},\ and\ \citenamefont {Sahashi}}]{shiokawa2011}%
  \BibitemOpen
  \bibfield  {author} {\bibinfo {author} {\bibfnamefont {Y.}~\bibnamefont
  {Shiokawa}}, \bibinfo {author} {\bibfnamefont {M.}~\bibnamefont {Shiota}},
  \bibinfo {author} {\bibfnamefont {Y.}~\bibnamefont {Watanabe}}, \bibinfo
  {author} {\bibfnamefont {T.}~\bibnamefont {Otsuka}}, \bibinfo {author}
  {\bibfnamefont {M.}~\bibnamefont {Doi}}, \ and\ \bibinfo {author}
  {\bibfnamefont {M.}~\bibnamefont {Sahashi}},\ }\href {\doibase
  10.1109/TMAG.2011.2157110} {\bibfield  {journal} {\bibinfo  {journal} {IEEE
  Trans. Magn.}\ }\textbf {\bibinfo {volume} {47}},\ \bibinfo {pages} {3470 }
  (\bibinfo {year} {2011})}\BibitemShut {NoStop}%
\bibitem [{\citenamefont {Ne\v{c}as}\ and\ \citenamefont
  {Klapetek}(2011)}]{necas2011}%
  \BibitemOpen
  \bibfield  {author} {\bibinfo {author} {\bibfnamefont {D.}~\bibnamefont
  {Ne\v{c}as}}\ and\ \bibinfo {author} {\bibfnamefont {P.}~\bibnamefont
  {Klapetek}},\ }\href {\doibase 10.2478/s11534-011-0096-2} {\bibfield
  {journal} {\bibinfo  {journal} {Cent. Eur. J. Phys.}\ }\textbf {\bibinfo
  {volume} {10}},\ \bibinfo {pages} {181} (\bibinfo {year} {2011})}\BibitemShut
  {NoStop}%
\bibitem [{\citenamefont {Kishi}\ \emph {et~al.}(2012)\citenamefont {Kishi},
  \citenamefont {Shiokawa}, \citenamefont {Watanabe}, \citenamefont {Zheng},\
  and\ \citenamefont {Sahashi}}]{kishi2012}%
  \BibitemOpen
  \bibfield  {author} {\bibinfo {author} {\bibfnamefont {K.}~\bibnamefont
  {Kishi}}, \bibinfo {author} {\bibfnamefont {Y.}~\bibnamefont {Shiokawa}},
  \bibinfo {author} {\bibfnamefont {H.}~\bibnamefont {Watanabe}}, \bibinfo
  {author} {\bibfnamefont {Z.}~\bibnamefont {Zheng}}, \ and\ \bibinfo {author}
  {\bibfnamefont {M.}~\bibnamefont {Sahashi}}\ }, presented at {{IEEE} International Magnetics Conference, Vancouver, British Columbia, Canada},\ \bibinfo {year}
  {2012} (unpublished)\BibitemShut {NoStop}%
\bibitem [{\citenamefont {Brinkman}, \citenamefont {Dynes},\ and\ \citenamefont
  {Rowell}(1970)}]{brinkman1970}%
  \BibitemOpen
  \bibfield  {author} {\bibinfo {author} {\bibfnamefont {W.}~\bibnamefont
  {Brinkman}}, \bibinfo {author} {\bibfnamefont {R.}~\bibnamefont {Dynes}}, \
  and\ \bibinfo {author} {\bibfnamefont {J.}~\bibnamefont {Rowell}},\
  }\href@noop {} {\bibfield  {journal} {\bibinfo  {journal} {J. Appl. Phys.}\
  }\textbf {\bibinfo {volume} {41}},\ \bibinfo {pages} {1915} (\bibinfo {year}
  {1970})}\BibitemShut {NoStop}%
\bibitem [{\citenamefont {Lu}\ \emph {et~al.}(1995)\citenamefont {Lu},
  \citenamefont {Newhouse}, \citenamefont {Dieckmann},\ and\ \citenamefont
  {Xue}}]{lu1995-a}%
  \BibitemOpen
  \bibfield  {author} {\bibinfo {author} {\bibfnamefont {F.}~\bibnamefont
  {Lu}}, \bibinfo {author} {\bibfnamefont {M.~L.}\ \bibnamefont {Newhouse}},
  \bibinfo {author} {\bibfnamefont {R.}~\bibnamefont {Dieckmann}}, \ and\
  \bibinfo {author} {\bibfnamefont {J.}~\bibnamefont {Xue}},\ }\href {\doibase
  10.1016/0167-2738(94)00147-K} {\bibfield  {journal} {\bibinfo  {journal}
  {Solid State Ionics}\ }\textbf {\bibinfo {volume} {75}},\ \bibinfo {pages}
  {187} (\bibinfo {year} {1995})}\BibitemShut {NoStop}%
\bibitem [{\citenamefont {Celinski}\ \emph {et~al.}(1990)\citenamefont
  {Celinski}, \citenamefont {Heinrich}, \citenamefont {Cochran}, \citenamefont
  {Muir}, \citenamefont {Arrott},\ and\ \citenamefont
  {Kirschner}}]{celinski1990}%
  \BibitemOpen
  \bibfield  {author} {\bibinfo {author} {\bibfnamefont {Z.}~\bibnamefont
  {Celinski}}, \bibinfo {author} {\bibfnamefont {B.}~\bibnamefont {Heinrich}},
  \bibinfo {author} {\bibfnamefont {J.~F.}\ \bibnamefont {Cochran}}, \bibinfo
  {author} {\bibfnamefont {W.~B.}\ \bibnamefont {Muir}}, \bibinfo {author}
  {\bibfnamefont {A.~S.}\ \bibnamefont {Arrott}}, \ and\ \bibinfo {author}
  {\bibfnamefont {J.}~\bibnamefont {Kirschner}},\ }\href {\doibase
  10.1103/PhysRevLett.65.1156} {\bibfield  {journal} {\bibinfo  {journal}
  {Phys. Rev. Lett.}\ }\textbf {\bibinfo {volume} {65}},\ \bibinfo {pages}
  {1156} (\bibinfo {year} {1990})}\BibitemShut {NoStop}%
\bibitem [{\citenamefont {Crangle}\ and\ \citenamefont
  {Scott}(1965)}]{crangle1965}%
  \BibitemOpen
  \bibfield  {author} {\bibinfo {author} {\bibfnamefont {J.}~\bibnamefont
  {Crangle}}\ and\ \bibinfo {author} {\bibfnamefont {W.~R.}\ \bibnamefont
  {Scott}},\ }\href {\doibase doi:10.1063/1.1714264} {\bibfield  {journal}
  {\bibinfo  {journal} {J. Appl. Phys.}\ }\textbf {\bibinfo {volume} {36}},\
  \bibinfo {pages} {921} (\bibinfo {year} {1965})}\BibitemShut {NoStop}%
\bibitem [{\citenamefont {Doudin}\ and\ \citenamefont
  {Viret}(2008)}]{doudin2008}%
  \BibitemOpen
  \bibfield  {author} {\bibinfo {author} {\bibfnamefont {B.}~\bibnamefont
  {Doudin}}\ and\ \bibinfo {author} {\bibfnamefont {M.}~\bibnamefont {Viret}},\
  }\href {\doibase 10.1088/0953-8984/20/8/083201} {\bibfield  {journal}
  {\bibinfo  {journal} {J. Phys.: Condens. Matter}\ }\textbf {\bibinfo {volume}
  {20}},\ \bibinfo {pages} {083201} (\bibinfo {year} {2008})}\BibitemShut
  {NoStop}%
\bibitem [{\citenamefont {Geissler}\ \emph {et~al.}(2001)\citenamefont
  {Geissler}, \citenamefont {Goering}, \citenamefont {Justen}, \citenamefont
  {Weigand}, \citenamefont {Sch\"utz}, \citenamefont {Langer}, \citenamefont
  {Schmitz}, \citenamefont {Maletta},\ and\ \citenamefont
  {Mattheis}}]{geissler2001}%
  \BibitemOpen
  \bibfield  {author} {\bibinfo {author} {\bibfnamefont {J.}~\bibnamefont
  {Geissler}}, \bibinfo {author} {\bibfnamefont {E.}~\bibnamefont {Goering}},
  \bibinfo {author} {\bibfnamefont {M.}~\bibnamefont {Justen}}, \bibinfo
  {author} {\bibfnamefont {F.}~\bibnamefont {Weigand}}, \bibinfo {author}
  {\bibfnamefont {G.}~\bibnamefont {Sch\"utz}}, \bibinfo {author}
  {\bibfnamefont {J.}~\bibnamefont {Langer}}, \bibinfo {author} {\bibfnamefont
  {D.}~\bibnamefont {Schmitz}}, \bibinfo {author} {\bibfnamefont
  {H.}~\bibnamefont {Maletta}}, \ and\ \bibinfo {author} {\bibfnamefont
  {R.}~\bibnamefont {Mattheis}},\ }\href {\doibase 10.1103/PhysRevB.65.020405}
  {\bibfield  {journal} {\bibinfo  {journal} {Phys. Rev. B}\ }\textbf {\bibinfo
  {volume} {65}},\ \bibinfo {pages} {020405} (\bibinfo {year}
  {2001})}\BibitemShut {NoStop}%
\bibitem [{\citenamefont {Suzuki}\ \emph {et~al.}(2005)\citenamefont {Suzuki},
  \citenamefont {Muraoka}, \citenamefont {Inaba}, \citenamefont {Miyagawa},
  \citenamefont {Kawamura}, \citenamefont {Shimatsu}, \citenamefont {Maruyama},
  \citenamefont {Ishimatsu}, \citenamefont {Isohama},\ and\ \citenamefont
  {Sonobe}}]{suzuki2005}%
  \BibitemOpen
  \bibfield  {author} {\bibinfo {author} {\bibfnamefont {M.}~\bibnamefont
  {Suzuki}}, \bibinfo {author} {\bibfnamefont {H.}~\bibnamefont {Muraoka}},
  \bibinfo {author} {\bibfnamefont {Y.}~\bibnamefont {Inaba}}, \bibinfo
  {author} {\bibfnamefont {H.}~\bibnamefont {Miyagawa}}, \bibinfo {author}
  {\bibfnamefont {N.}~\bibnamefont {Kawamura}}, \bibinfo {author}
  {\bibfnamefont {T.}~\bibnamefont {Shimatsu}}, \bibinfo {author}
  {\bibfnamefont {H.}~\bibnamefont {Maruyama}}, \bibinfo {author}
  {\bibfnamefont {N.}~\bibnamefont {Ishimatsu}}, \bibinfo {author}
  {\bibfnamefont {Y.}~\bibnamefont {Isohama}}, \ and\ \bibinfo {author}
  {\bibfnamefont {Y.}~\bibnamefont {Sonobe}},\ }\href {\doibase
  10.1103/PhysRevB.72.054430} {\bibfield  {journal} {\bibinfo  {journal} {Phys.
  Rev. B}\ }\textbf {\bibinfo {volume} {72}},\ \bibinfo {pages} {054430}
  (\bibinfo {year} {2005})}\BibitemShut {NoStop}%
\bibitem [{\citenamefont {Sahashi}\ \emph {et~al.}(2007)\citenamefont
  {Sahashi}, \citenamefont {Sawada}, \citenamefont {Endo}, \citenamefont
  {Doi},\ and\ \citenamefont {Hasegawa}}]{sahashi2007}%
  \BibitemOpen
  \bibfield  {author} {\bibinfo {author} {\bibfnamefont {M.}~\bibnamefont
  {Sahashi}}, \bibinfo {author} {\bibfnamefont {K.}~\bibnamefont {Sawada}},
  \bibinfo {author} {\bibfnamefont {H.}~\bibnamefont {Endo}}, \bibinfo {author}
  {\bibfnamefont {M.}~\bibnamefont {Doi}}, \ and\ \bibinfo {author}
  {\bibfnamefont {N.}~\bibnamefont {Hasegawa}},\ }\href {\doibase
  10.1109/TMAG.2007.894012} {\bibfield  {journal} {\bibinfo  {journal} {IEEE
  Trans. Magn.}\ }\textbf {\bibinfo {volume} {43}},\ \bibinfo {pages} {3668}
  (\bibinfo {year} {2007})}\BibitemShut {NoStop}%
\bibitem [{\citenamefont {Slonczewski}(1995)}]{slonczewski1995}%
  \BibitemOpen
  \bibfield  {author} {\bibinfo {author} {\bibfnamefont {J.}~\bibnamefont
  {Slonczewski}},\ }\href@noop {} {\bibfield  {journal} {\bibinfo  {journal}
  {J. Magn. Magn. Mater.}\ }\textbf {\bibinfo {volume} {150}},\ \bibinfo
  {pages} {13} (\bibinfo {year} {1995})}\BibitemShut {NoStop}%
\bibitem [{\citenamefont {Fukuzawa}\ \emph {et~al.}(2002)\citenamefont
  {Fukuzawa}, \citenamefont {Koi}, \citenamefont {Tomita}, \citenamefont
  {Fuke}, \citenamefont {Iwasaki},\ and\ \citenamefont
  {Sahashi}}]{fukuzawa2002}%
  \BibitemOpen
  \bibfield  {author} {\bibinfo {author} {\bibfnamefont {H.}~\bibnamefont
  {Fukuzawa}}, \bibinfo {author} {\bibfnamefont {K.}~\bibnamefont {Koi}},
  \bibinfo {author} {\bibfnamefont {H.}~\bibnamefont {Tomita}}, \bibinfo
  {author} {\bibfnamefont {H.~N.}\ \bibnamefont {Fuke}}, \bibinfo {author}
  {\bibfnamefont {H.}~\bibnamefont {Iwasaki}}, \ and\ \bibinfo {author}
  {\bibfnamefont {M.}~\bibnamefont {Sahashi}},\ }\href {\doibase
  http://dx.doi.org/10.1063/1.1471364} {\bibfield  {journal} {\bibinfo
  {journal} {J. Appl. Phys.}\ }\textbf {\bibinfo {volume} {91}},\ \bibinfo
  {pages} {6684} (\bibinfo {year} {2002})}\BibitemShut {NoStop}%
\end{thebibliography}
\end{document}